\documentclass[sigconf]{acmart}

\AtBeginDocument{%
  \providecommand\BibTeX{{%
    \normalfont B\kern-0.5em{\scshape i\kern-0.25em b}\kern-0.8em\TeX}}}

\setcopyright{acmcopyright}

\copyrightyear{2020}

\acmYear{2020}

\acmDOI{xx.xxxx/xxxxxxxxx.xxxxxxx}

\acmConference[epiDAMIK 2020]{3rd epiDAMIK SIGKDD International Workshop on Epidemiology meets Data Mining and Knowledge Discovery}{Aug 24, 2020}{San Diego, CA}

\acmBooktitle{epiDAMIK 2020: 3rd epiDAMIK SIGKDD International Workshop on Epidemiology meets Data Mining and Knowledge Discovery}

\acmPrice{15.00}

\acmISBN{978-1-xxxx-XXXX-X}

\usepackage{graphicx}
\usepackage{booktabs}
\usepackage{multirow}
\usepackage{subfig}
\usepackage[inline]{enumitem}

\newlist{inlinelist}{enumerate*}{1}
\setlist*[inlinelist,1]{%
  label=(\roman*),
}
\begin{document}

\title{A Canine Census to Influence Public Policy}

\author{Matias Apa, Maria Cecilia Faini}
\email{{matias_apa, mcfaini}@yahoo.com.ar}
\affiliation{%
  \institution{Facultad de Ciencias Veterinarias \\ Universidad Nacional de Rosario}
  \city{Casilda}
   \state{Santa Fe}
  \country{Argentina}
}

\author{Mohammad Aliannejadi}
\email{m.aliannejadi@uva.nl}
\affiliation{%
  \institution{University of Amsterdam}
  \city{Amsterdam}
   \country{The Netherlands}
}

\author{Maria Soledad Pera}
\email{solepera@boisestate.edu}
\orcid{0000-0002-2008-9204}
\affiliation{%
  \institution{People and Information Research Team (PIReT), Boise State University}
  \city{Boise}
  \state{Idaho}
  \country{USA}
}

\renewcommand{\shortauthors}{Apa, et al.}

\begin{abstract}
The potential threat that domestic animals pose to the health of human populations tends to be overlooked. We posit that positive steps forward can be made in this area, via suitable state-wide public policy. In this paper, we describe the data collection process that took place in Casilda (a city in Argentina), in the context of a canine census. We outline preliminary findings emerging from the data, based on a number of perspectives, along with implications of these findings in terms of informing public policy.
\end{abstract}


%\ccsdesc[500]{Human-centered computing~Information visualization}


\maketitle

%===============================================
\section{Introduction}
Using the epidemiology of urban health as a lens, we can study the environment and context of a region to understand (i) the ties and relationships of species among themselves and with the environment, (ii) the complexity of the urban context, and (iii) the consequences that result from these complex interactions and the social determinants of health~\cite{spiegel2015language}. Ecosystems and human health are deep-rooted on  biological processes that are socially defined. The fact that social mandates influence health-related determinations posits a dialectical perspective to explore ``social-biological'' and ``society-nature'' interactions, both of which contribute towards the phenomenology of health. The transformation patterns observed between society and the environment are continually evolving; yet, social determinations are hierarchically imposed and are the ones that most prominently prevail in nature~\cite{breilh2010epidemiologia}.

Domestic animals are a clear example of social constructs prevailing over biological ones. In their case, dynamics within the population are defined by social norms and standards, as well as political and cultural practices. The number of animals in a given region depends on the availability of resources (food, water, shelter) and human acceptance of the particular population. This is the reason why canine ecology is deeply interconnected with human-related activities~\cite{OIE2010}. Ongoing development of regions results in changes in habits and behavior of their inhabitants, such as the increase in the number of companion animals that are now part of households, especially dogs and cats. The bond between humans and companion animals has both positive and negative effects on health. Examples of the latter are zoonoses, animal bites, and pollution. It is worth noting that all the concerns become more critical when these animals have access to public roads~\cite{loza2014caracterizacion}. 

Policy and campaigning messages to promote a healthy human-animal coexistence depend on a better understanding of the companion animals' social placement and dynamics in a region. This can be achieved by collecting representative data and analyzing the demographic characteristics of animal populations, local traits, and natural human-animal interactions~\cite{bovisio2004caracteristicas}. Our study focuses on the data collected in Casilda (Santa Fe), Argentina. For decades, the local community has demanded that the city council address concerns related to dog ownership and welfare. In fact, the city council introduced an ordinance concerning the canine census in 2008~\cite{Ordenanza}. However, there has been a long delay before we  conducted the first census in 2018 due to a lack of study protocols.  

In this paper, we present the results of Casilda's first-ever domestic canine census. Doing so involved an interdisciplinary group of faculty and students in epidemiology, public health, ethics, and legislation for veterinary sciences, statistics, and computer science, who were mindful of the economic and political constraints inherent of the region and designed a protocol for data collection and conducted the associated analysis. %, . designed and deployed a canine census, in order to generate necessary knowledge to depict Casilda's domestic canine population, its connection with the social context, the environment, and the community. To respond to the needs and constraints inherent to the current politic,
The main objectives guiding our work include: 
\begin{itemize}[leftmargin=*]
    \item Providing an exhaustive description of Casilda's domestic canine population. To do so, we conducted an empirical exploration of census data; this revealed scientific indicators that allow assessment of the quality of life of canines, along with the quality of their interactions with humans.
    \item Identifying new problems that emerge from the surveyed canine population can be addressed by new public policies, thus responding to the demands of the region and its inhabitants \cite{Apa19}.
\end{itemize}

Our work lays the grounds for future work in this area by introducing the necessary protocols that could be used in similar studies. Moreover, it aids the local government in making informed decisions in response to the existing canine-human problems.

%In the rest of this manuscript, we first describe the data collection protocol (Section \ref{sec:Protocol}). In Section \ref{sec:Analysis}, we present findings emerging from census data; in addition to possible implications of such findings on public policy. We offer concluding remarks in Section \ref{sec:Conclusions}.

%===============================================
\section{Protocol for Data Collection}
\label{sec:Protocol}
\begin{figure}[h]
\centering
\includegraphics[width=0.375\textwidth]{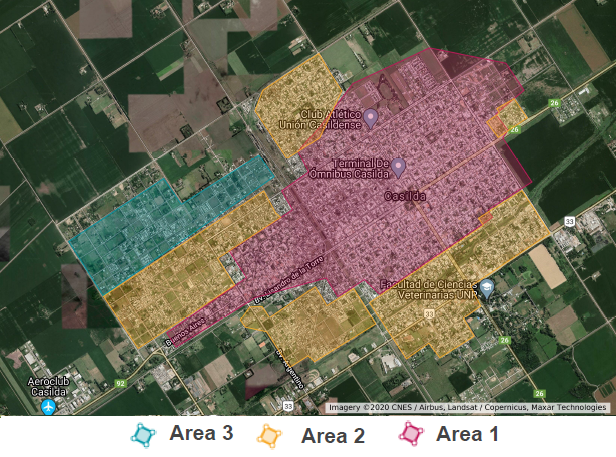}
\caption{Geographical areas considered in the census.}
\label{fig:AreaMap}
\end{figure}

We collected census data and conducted systematic probabilistic sampling by areas. In establishing these areas, we considered different traits (social, environmental, and economic) that characterize Casilda (population 37,441). This resulted in the 3 geographical areas (Figure \ref{fig:AreaMap}): \textbf{Area 1} (5.22$km^{2}$), upper/upper middle class; \textbf{Area 2} (3.82$km^{2}$), middle class; \textbf{Area 3} (1.50$km^{2}$), working/lower class. 

Data collection took place in June 2018; involving a team of 80 students and 18 faculty in the Epidemiology Department at Universidad Nacional de Rosario \footnote{Training for data collection is part of the curriculum for one of the epidemiology-related classes offered at Universidad Nacional de Rosario \cite{FainiCongreso18}.}. The surveyed area included 60 blocks (1,189 households that resulted in 486 voluntary responses). The team reported a general low predisposition on behalf of household occupants in taking part in the census. This rendered the sample insufficient for statistical inference. To address this limitation, the team sub-sampled 125 new households to survey \cite{azorin1994metodos}. 

For data collection, the team created a dynamic form with response-depended questions using Google Forms. The questionnaire included 26 questions (Table~\ref{tab:CensusQuestionnaire}), some closed-ended and others multiple-choice; grouped by data related to \textit{households}, \textit{household occupants}, \textit{canines}, \textit{responsible ownership}, and \textit{general}. Google Forms was chosen as it is a free tool that eases immediate digitization of the collected data, reducing operational costs and the use of paper -- these are constraints that influenced data collection decisions, given that resources at public universities in Argentina are scarce. 

\begin{table}[]
\centering
\caption{Questionnaire used for data collection purposes.}
\label{tab:CensusQuestionnaire}
\resizebox{\linewidth}{!}{%
\begin{tabular}{@{}ccl@{}}
\toprule
\textbf{Type} &
  \textbf{ID} &
  \multicolumn{1}{c}{\textbf{Question}} \\ \midrule
\multirow{4}{*}{Household} &
  1 &
  Area \\
 &
  2 &
  Address \\
 &
  3 &
  Household type \\
 &
  4 &
  Services (e.g., gas, water, etc.) \\
\multirow{2}{*}{\begin{tabular}[c]{@{}c@{}}Household \\ occupants\end{tabular}} &
  5 &
  Are they in and willing to answer questionnaire? \\
 &
  6 &
  How many people live in the household? \\
\multirow{6}{*}{\begin{tabular}[c]{@{}c@{}}Canines\\ (repeated \\ for each \\ dog in \\ household)\end{tabular}} &
  7 &
  Breed \\
 &
  8 &
  Gender \\
 &
  9 &
  Age \\
 &
  10 &
  Size \\
 &
  11 &
  Origin (e.g., adopted,  found, etc.) \\
 &
  12 &
  Sterilized? \\
\multirow{9}{*}{\begin{tabular}[c]{@{}c@{}}Responsible\\ ownership\end{tabular}} &
  13 &
  Where did the sterilization take place? \\
 &
  14 &
  If not, why not? \\
 &
  15 &
  Where does your dog live? (patio, indoors, etc.) \\
 &
  16 &
  How often in your dog on public roads? \\
 &
  17 &
  If veterinary services are required, where do you go? \\
 &
  18 &
  How often do you deworm your dog? (internally) \\
 &
  19 &
  How often do you deworm your dog (externally) \\
 &
  20 &
  \begin{tabular}[c]{@{}l@{}}In the last year, have you vaccinated your dog for rabies? Where?\end{tabular} \\
 &
  21 &
  \begin{tabular}[c]{@{}l@{}}In the last year, have you vaccinated your dog for Leptospirosis? Where?\end{tabular} \\
\multirow{5}{*}{\begin{tabular}[c]{@{}c@{}}General\end{tabular}} &
  22 &
  Are there any other animals in the household. Elaborate. \\
 &
  23 &
  \begin{tabular}[c]{@{}l@{}}In the last year, have you experienced any of the following: \\ bites, dog involved in accident, chasing bicycles \\ and/or people walking, saw canine mistreatment, etc.\end{tabular} \\
 &
  24 &
  Do you know your neighbourhood's health center? \\
 &
  25 &
  Regarding your neighbourhood's health center \\
 &
  26 &
  For your own health-related matters, where do you go? \\ \bottomrule
\end{tabular}%
}
\end{table}

%===============================================
\section{Analysis and Discussion}
\label{sec:Analysis}
Below we summarize general observations that emerged from collected data; these are meant to offer context of the geographical area and human and canine populations considered in the census. Thereafter, we present detailed findings from census data, along with their implications for public policy.

\subsection{A General Description of the Population}
\label{subsec:General}

Based on collected responses, we analyze the data of 841 dogs, uniformly distributed across gender. We summarize sanitary conditions %(i.e., deworming and vaccination) 
and sterilization in Table \ref{tab:overviewCaninePopulation}. Other insights include:
\begin{itemize}[leftmargin=*]
    \item \textbf{Breed}: 33\%  were pure-breeds, the rest mongrels.
    \item \textbf{Size}: close to 50\% were small breeds (e.g., Beagle, Poodle Toy), 33 \% medium (e.g., French bulldog), and the remaining, larger breeds (e.g., Golden retriever).
    \item \textbf{Origin}: 74\% were either adopted or found, 20\% were purchased, and for the remaining ones, survey respondents did not recall. 
   \item \textbf{Age}: 498 were adults between 1 and 7 years old, 154 were puppies (i.e., less than 12 months), and the remaining 189 were seniors (i.e., 8 years or more). 
   \item \textbf{Area}: 233 in Area 1, 401 in Area 2, and 207 in Area 3.
   \item \textbf{Inference from sampling}: 13,557 dogs in households,  4,863 strays \cite{FainiCongreso18}.
\end{itemize}

\begin{table}[h]
\centering
\caption{Overview of surveyed canine population}
\label{tab:overviewCaninePopulation}
\resizebox{0.8\linewidth}{!}{%
\begin{tabular}{@{}lccc@{}}
\toprule
\textbf{Canines}          & \multicolumn{1}{c}{\textbf{Total}}  & \multicolumn{1}{c}{\textbf{Male}} & \multicolumn{1}{c}{\textbf{Female}} \\ \midrule
Surveyed                  & 841                                 & 422                               & 419 \\
Sterilized                & 318                                 & 67                                & 251 \\
Internal deworming   & 692                                 & 344                               & 348                      \\
External deworming   & 728                                 & 364                               & 364                      \\
Rabies vaccination        & 440                                 & 219                               & 221                      \\
Leptospirosis vaccination & 299                                 & 137                               & 162                      \\ \bottomrule
\end{tabular}%
}
\end{table}

\subsection{Findings and Implications}
\label{subsec:Details}
To further characterize Casilda's domestic canine population, and more importantly, identify issues directly related to this population, we further examined census responses from various perspectives. 

\subsubsection{Sterilization}
\begin{figure}[]
\centering
\includegraphics[width=0.75\columnwidth]{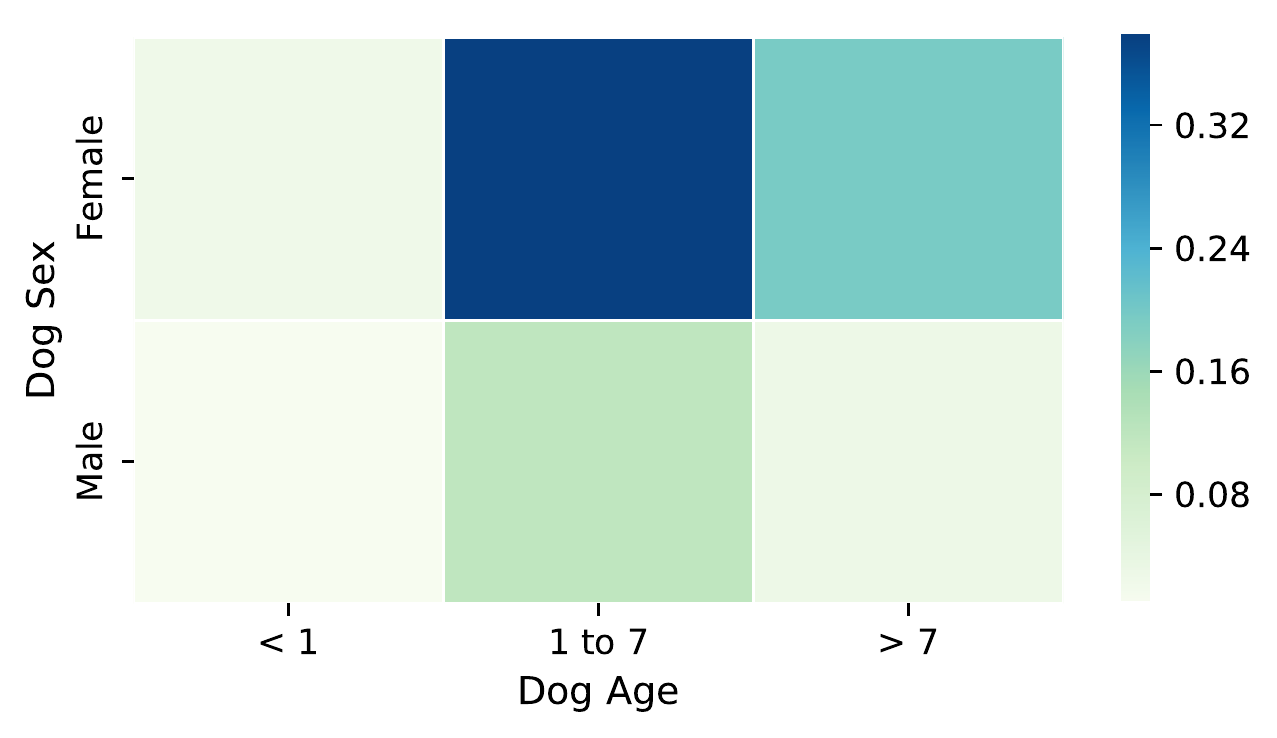}
\caption{Correlation with respect to sterilization.}
\label{fig:Sterelization}
\end{figure}

We see a statistical significant correlation between gender and age, when it comes to sterilization (Chi-square: 24.85; p-value= 5.38e-05).
As reported in Table \ref{tab:overviewCaninePopulation}, close to 40\% of domestic canines have been sterilized; for the most part, females. It is also apparent from Figure~\ref{fig:Sterelization} that sterilization rarely occurs on canines less than 12 months old; the majority of sterilizations happening on adult specimens (i.e., aged  1 to 7).
As for why owners bypass sterilization (question 14 in Table \ref{tab:CensusQuestionnaire}), close to 30\% \textit{``do not think it is necessary''} and 3.4\% \textit{``disagrees with the premise of sterilization''}. It is of note that 13\% of the owners \textit{``plan sterilization in the future''} and 1.3\% have yet to do so \textit{``due to economic impediments''}. 

Female sterilization is a positive discovery, especially when considering that it occurs at an age range that correlates with the highest fertility peaks. Unfortunately, lack of sterilization in males counteracts intended population control. Further, the high proportion of unsterilized males is a definite concern that must be addressed. Their social behavior entails wandering and territoriality, often resulting in dog fights, bites of people, the transmission of diseases, and traffic accidents. Owners' views against sterilization reflect that population control policy must be thought of as a comprehensive scheme. The system must ensure the economic and geographical accessibility to an operating room. It must also include educational strategies that raise awareness of the negative consequences of non-sterilization.

%: X-axis age (in years); Y-axis gender.
\subsubsection{Sanitary Conditions}
We examine the degree of influence, or lack thereof, that the number of dogs per household has on traits related to responsible ownership practises (questions 18-21 in Table \ref{tab:CensusQuestionnaire}). We find that \textit{deworming} (internal or external) and \textit{vaccinations} for \textit{rabies} are not conditioned by the number of dogs in a household. There is a statistical significant correlation between \textit{vaccination} for \textit{Leptospirosis} and number of dogs per households, where more dogs implies a higher likelihood of overlooking this type of vaccination (ANOVA, p-value = 0.001; Figure \ref{fig:lepto_household}).

These results show the broad access that the local population has to the rabies vaccine. In Argentina, Law No. 22953 establishes this % rabies
vaccine as mandatory. The city sponsors free %rabies 
vaccination campaigns, together with the application of dewormers. Further, deworming is a low-cost procedure when performed at private veterinary clinics. On the other hand, the Leptospirosis vaccine is not part of sponsored campaigns, and Leptospirosis vaccination at private clinics is very expensive. Therefore, state policy responding to this concern should include targeted campaigns on high-risk areas. 

% \begin{figure}[]
% \centering

% \includegraphics[width=0.30\textwidth]{Figuras/lepto_household.png}
% \caption{Leptospirosis vaccination vs. dogs per household.}
% \label{fig:lepto_household}
% \end{figure}

\subsubsection{Socio-economic Influence}
\begin{figure}[]
\centering
\subfloat[Dogs per household]{\includegraphics[height=44mm]{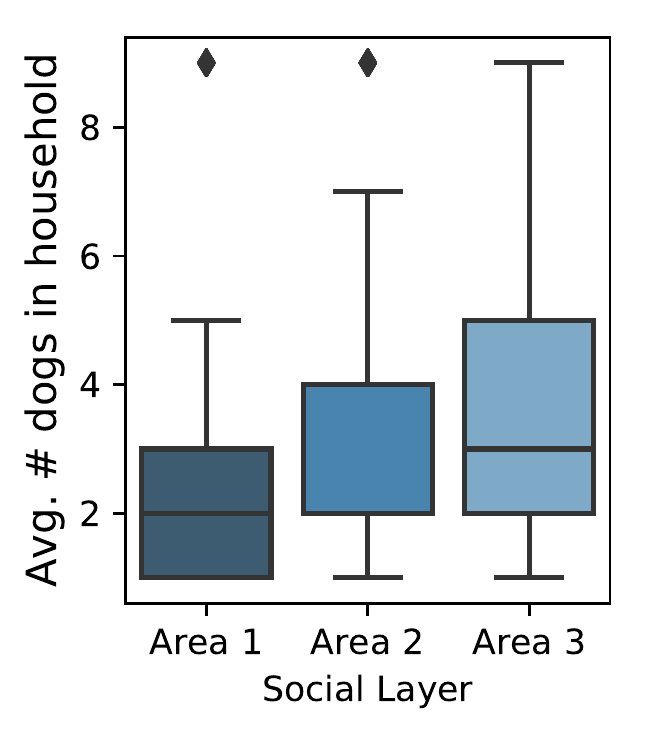}\label{fig:Dogs_area}}
~
\subfloat[Vaccination per household]{\includegraphics[height=44mm]{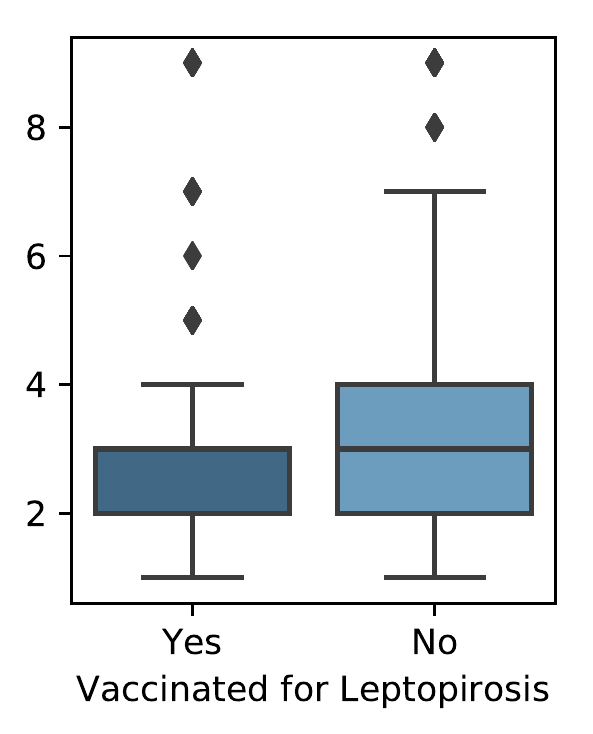}\label{fig:lepto_household}}
% \includegraphics[width=0.30\textwidth]{Figuras/Dogs_by_area.png}
% \caption{Average number of dogs per household, distributed by areas under study.}
\label{fig:per_area}
\caption{Average number of dogs per household distributed by areas (a) and the ones vaccinated for Leptospirosis (b).}
\end{figure}

\begin{figure}[]
\centering
\includegraphics[width=0.8\columnwidth]{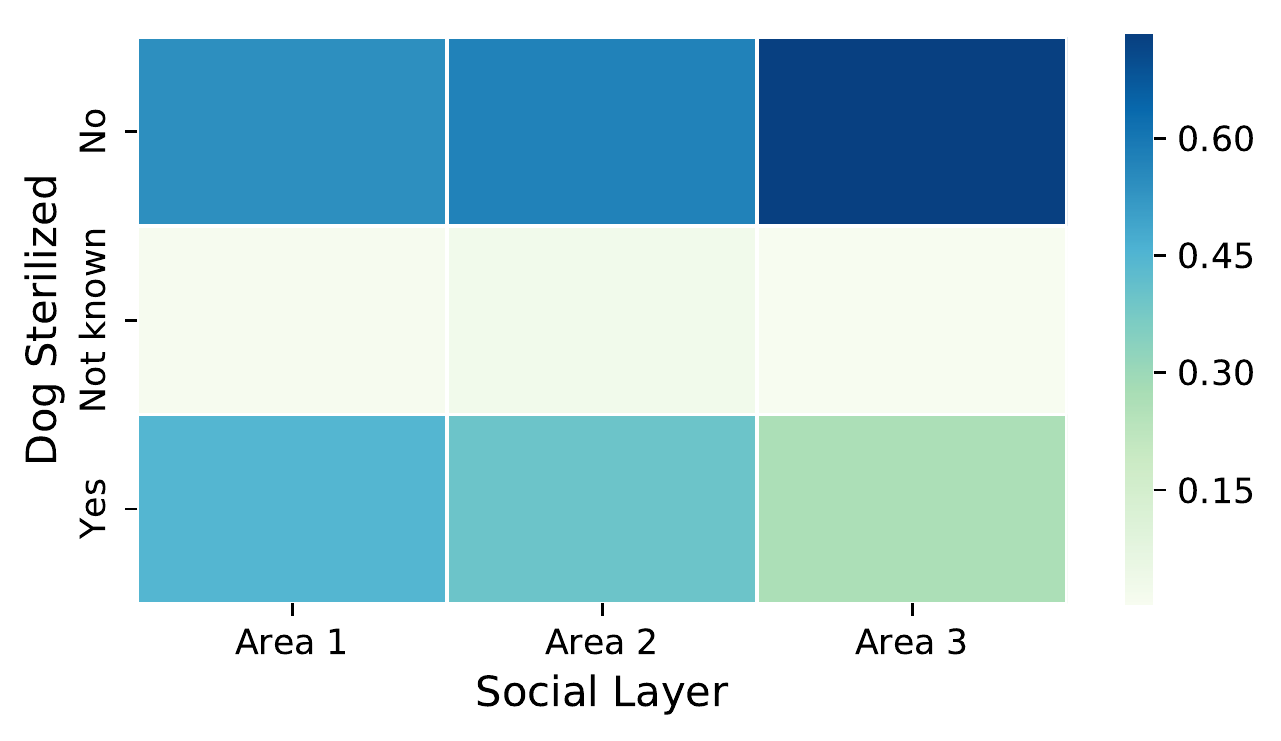}
\caption{Canine sterilization across areas. ``Not known'' refer to cases when the dog owner was not present to answer.}
% In some households, domestic workers where the ones providing answers to the census, which resulted in "Not known" responses.}
\label{fig:Sterelization_area}
\end{figure}

\begin{figure}[]
\centering
\includegraphics[width=0.875\columnwidth]{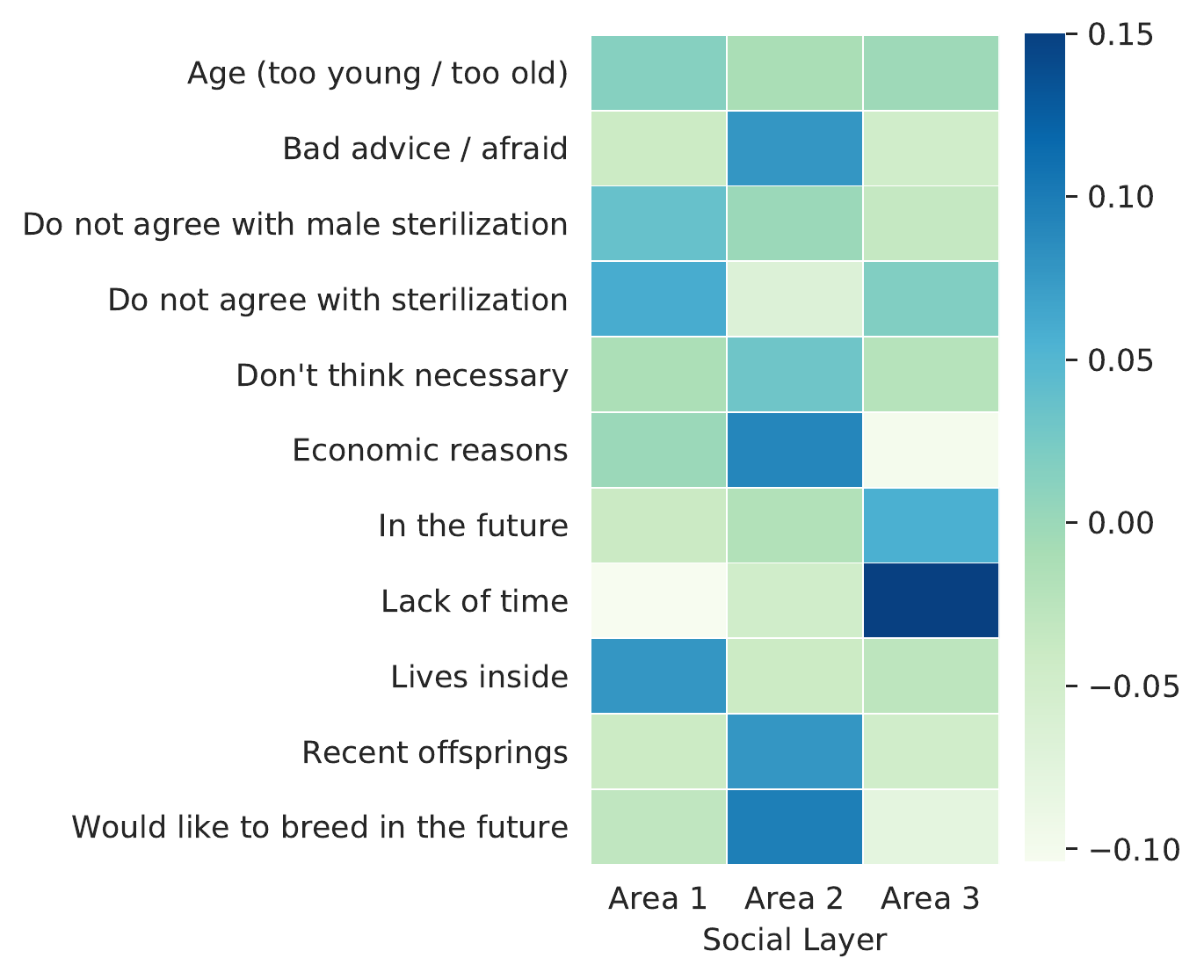}
\vspace{-0.1in}
\caption{Pearson's correlation for area with respect to reasons why household owners chose to avoid sterilization.}
\vspace{-0.2in}
\label{fig:NoSterelization_area}
\end{figure}

When using socio-economic factors as lenses to drive exploration, census data reveals a correlation between geographic areas and number of dogs per households (Figure \ref{fig:Dogs_area} Chi-square: 22.87; p-value= 0.00013). Upon deeper inspection, we see that the highest percentage of unsterilized dogs come from households in Area 3, whereas most sterilized dogs come from households in Area 1 (Figure~\ref{fig:Sterelization_area}). Based on Pearson's correlation among area and reasons given by household owners to justify they do not favor sterilziation (Figure~\ref{fig:NoSterelization_area}), we see that the reason that yields the highest correlation for households in the least affluent area (i.e., Area 3) is \textit{``Lack of time''}, followed by \textit{``I will do so in the future''}. On the other spectrum, households in more affluent areas (i.e., Area 1 and Area 2) justify not sterilizing their dogs since they \textit{``Lives inside''} and \textit{``Would like to breed in the future''}, respectively.

The results above evidence the fact that low-income regions should be the focus of attention for public policy related to responsible pet ownership. In these regions, it is imperative to ensure economic and geographical assistance by performing State-sponsored (i.e., free) sterilization in peripheral neighborhoods.
Despite the fact that lack of time is not an impediment for sterilization in Areas 1 and 2, sterilization rates are not 100\% in these areas (Figure \ref{fig:Sterelization_area}); on Area 3, lack of time is the main issue hindering sterilization procedures. These findings could suggest that education and awareness campaigns would be more effective in Areas 1 and 2, whereas those mentioned earlier sponsored and geographically-targeted sterilization procedures could be more effective for Area 3.

%alternative: {Figuras/Nosterilization_area.png}

\subsubsection{Humans}
Canines with frequent access to public roads pose a risk to human health. %From responses to question 16 in Table \ref{tab:CensusQuestionnaire}, 
We identified 687 dogs that have access to public roads; 362 males, 325 females (question 16 in Table \ref{tab:CensusQuestionnaire}). 
As reported in Table \ref{tab:OutsideAccess}, only 50\% of these dogs are vaccinated for rabies--a low percentage when considering that this vaccination is mandatory. The percentage decreases even further for Leptospirosis ($\sim$ 30\%). Compared to vaccinations, the percentage of frequently-dewormed dogs with access to public roads is much higher ($\sim$ 75\%). 

The high proportion of dogs with access to public roads is a threat to public health. Because of unvaccinated dogs, the risk of exposure to diseases increases. Leptospirosis is an endemic zoonotic disease in Casilda. Thus actions by the State to address the low vaccination coverage are a must. Rabies-related concerns are much more worrisome:  given that in addition to being a lethal zoonosis, there is evidence of the circulation of this virus in Casilda, vaccinations rates reach  100\%. On the upside, the high proportion of dewormed dogs is positive for health care, as it prevents disease spread to other dogs and humans, which can be done via contaminated dog feces or ticks, to name a few.

\subsubsection{Overpopulation}
Dogs with access to public roads may cause an unexpected increase in canine populations: specimens that have not been sterilized, yet have access to public roads are bound to become a link in a chain of unplanned litters. As shown in Table \ref{tab:OutsideAccess}, 60\% of females with access to public roads are sterilized, a percentage that drastically decreases among males ($\sim$ 14\%). 

When campaigns fostering sterilization are not prominent, dog population growth rates remain high. Given the significant proportion of unsterilized females with access to public roads, compounded by the very high percentage of unsterilized males, breeding likelihood is high. That is why sterilization mechanism should be intensified, with a greater emphasis on males and social sectors with economic difficulties (i.e., low-income areas). These actions should be supplemented with an educational policy that emphasizes the importance of long-term behavior change regarding responsible pet ownership, specifically adopting new habits that foster health care for dogs and their environment.

\begin{table}[]
\centering
\caption{Canine population that has access (on its own, leashed and/or unleashed) to public roads (n=687), based on vaccination, deworming, and sterilization perspectives.}
\label{tab:OutsideAccess}
\resizebox{0.8\linewidth}{!}{%
\begin{tabular}{@{}llccc@{}}
\toprule
\multicolumn{2}{c}{\multirow{2}{*}{}}      & \multicolumn{3}{c}{\textbf{Canine Population}} \\ 
\multicolumn{2}{c}{}                       & Total       & Male      & Female      \\ \hline
\textbf{Vaccination}      & Rabies         &    0.52         & 0.49      & 0.54        \\
\multicolumn{1}{l}{}      & Leptospirosis  &      0.33       & 0.30      & 0.35        \\
\textbf{Deworming}   & Internal       &      0.74       & 0.72      & 0.76        \\
\multicolumn{1}{l}{}      & External       &     0.75        & 0.74      & 0.75        \\
\textbf{Sterilization} & ~   &  0.36        & 0.14      & 0.60         \\ \bottomrule
\end{tabular}%
}
\end{table}
%===============================================
\section{Conclusions}
\label{sec:Conclusions}
We have presented the analysis results we conducted on data collected in response to a domestic canine population census.  

Outcomes from our empirical exploration reveal representative traits of Casilda's canine population, which till now were unavailable.  We were also able to recognize potential risks originated from the population under study, mainly the transmission of zoonosis and uncontrolled breeding. At the same time, we identified geographic areas and social stratum that should be of primary concern to the city council when it comes to implementing immediate actions regarding sterilization, improvement of sanitary conditions, and education related to responsible pet ownership. This study serves as preliminary evidence on the importance of generating information on canine demography and its link with humans and the environment at the national level. An adapted version of the proposed data collection/analysis protocol -- based on lessons learned and limitations we observed -- could be included as part of the national population census, which takes place every ten years.

\begin{acks}
We appreciate Ion Madrazo Azpiazu's  feedback on data collection and Federico Abud's work on statistical inference.
\end{acks}

\bibliographystyle{ACM-Reference-Format}
\bibliography{Referencias}


\end{document}